# Imagining the Alien:
# Human Projections and Cognitive Limitations

_______________________________________________________________

## S. G. Djorgovski

*California Institute of Technology, Pasadena, CA 91125, USA*

ABSTRACT: Imagining what life on other planets, and intelligent life in particular, may be like is a long-running theme in human culture. It is a manifestation of the innate human curiosity about the Cosmos, and it has inspired numerous works of art and folklore, including whole literary and other media genres. It is a profound question, with philosophical and existential implications. There is also an obvious connection with religious beliefs, as gods and other superhuman beings were imagined in the heavens. Speculations about alien beings grew in time, and today, it is a scientific subject of astrobiology, and it is pursued through serious searches for life and intelligence in the universe. However, almost all imaginings of the alien map terrestrial life forms and human cultural, historical, and psychological phenomena to the putative aliens. This lack of individual and collective imagination may reflect our biological and cultural evolution, as our minds are formed through our experiences, perceptions of the world, and interactions with our terrestrial and human environments. As such, imagining aliens is mainly a cultural phenomenon and may reflect the intrinsic cognitive limitations of the human mind. Interestingly, we did create what is effectively an alien intelligence on this planet in the form of now rapidly evolving Artificial Intelligence (AI). As its capabilities grow, it may give us new insights into what extraterrestrial advanced intelligences may be like.

For thousands of years, humans have wondered about the inhabitants of the heavens. They had to be special to live there, so most religions assumed that is where their god(s) reside(s). The gods also conveniently looked and dressed just like their faithful at the time, had the same technologies, and the same lack of knowledge about nature and the physical universe. The hypothetical heavenly beings were imagined to be just like us, or very similar to us physically, and they also acted like us, with similar motivations. The delightful soap opera that was Greek mythology is an excellent example of such projections.

Philosopher scholars of classical Greece went beyond that. They postulated a plurality of worlds, many of which are to be inhabited by creatures unknown. After all, their gods lived on a mountaintop nearby. This probably started with Anaximander, followed by Democritus, Leucippus, Epicurus, Anaxarchus, and others[1]. Plutarch reports that Anaxarchus told Alexander the Great that there was an infinite number of worlds, "causing the latter to become dejected because he had not yet conquered even one". This theme of military conquest of other planets is alive and well in modern science fiction, although often Earth is that other planet.

The idea of the plurality of worlds was rejuvenated in the Copernican revolution, in particular by Giordano Bruno[2]. In his work *De l'infinito universo et mondi* (1584) he speculated about the infinity of worlds, populated by other living creatures. It is fair to say that his ideas were not well received by the church, which contributed to his untimely death.

It was natural that such ideas would be revived during the Enlightenment. Bernard Le Bovier de Fontenelle, in his popular science book *Entretiens sur la pluralité des mondes* (Conversations on the Plurality of Worlds) in 1686, speculated that the Moon and other planets are inhabited and that there are also planets orbiting other stars[3]. He may be considered a forerunner of the now blossoming discipline of astrobiology (or bioastronomy – it depends on whether you ask an astronomer or a biologist). Christiaan Huygens, in his book *Cosmotheoros* (1698), also speculated extravagantly on the existence of extraterrestrial life, which he imagined to be similar to that on Earth[4].

With the Industrial Revolution, there was a growth of the public interest in science, including worlds other than the Earth. One interesting episode is *The Great Moon Hoax of 1835*, a six-part series of articles published by the newspaper *The Sun*[5]. It reported on completely fictitious discoveries of life and civilization on the Moon. The discoveries were falsely attributed to Sir John Herschel, a well-known astronomer who was in South Africa at the time; he was initially amused by this once he found out. The articles caused a sensation. They claimed that the Moon was inhabited by humans, although with fur and bat-like wings, as well as animals that looked like their terrestrial counterparts with some modifications.

---

[1] Richard Sorabji and Michael Griffin, eds., *Simplicius: On Aristotle Physics 1.5-9 (Ancient Commentators on Aristotle)*, Reprint Edition (New York: Bloomsbury Academic, 2014)

[2] Russell Genet, *Giordano Bruno's cosmic hypothesis: The universe is infinite, evolving, and filled with planets, life and intelligence*, (Union Institute and University ProQuest Dissertations & Theses, 2003)

[3] Bernard le Bovier de Fontenelle, *Conversations on the Plurality of Worlds*, reprint, (University of California Press, 1990)

[4] Christiaan Huygens, *Cosmotheoros: Or Conjectures Concerning the Planetary Worlds, and Their Inhabitants* (Gale Ecco, Print Editions, 2018)

[5] Stephanie Hall, *Belief, Legend, and the Great Moon Hoax* (Library of Congress Blogs, 2014), available at https://blogs.loc.gov/folklife/2014/08/the-great-moon-hoax/

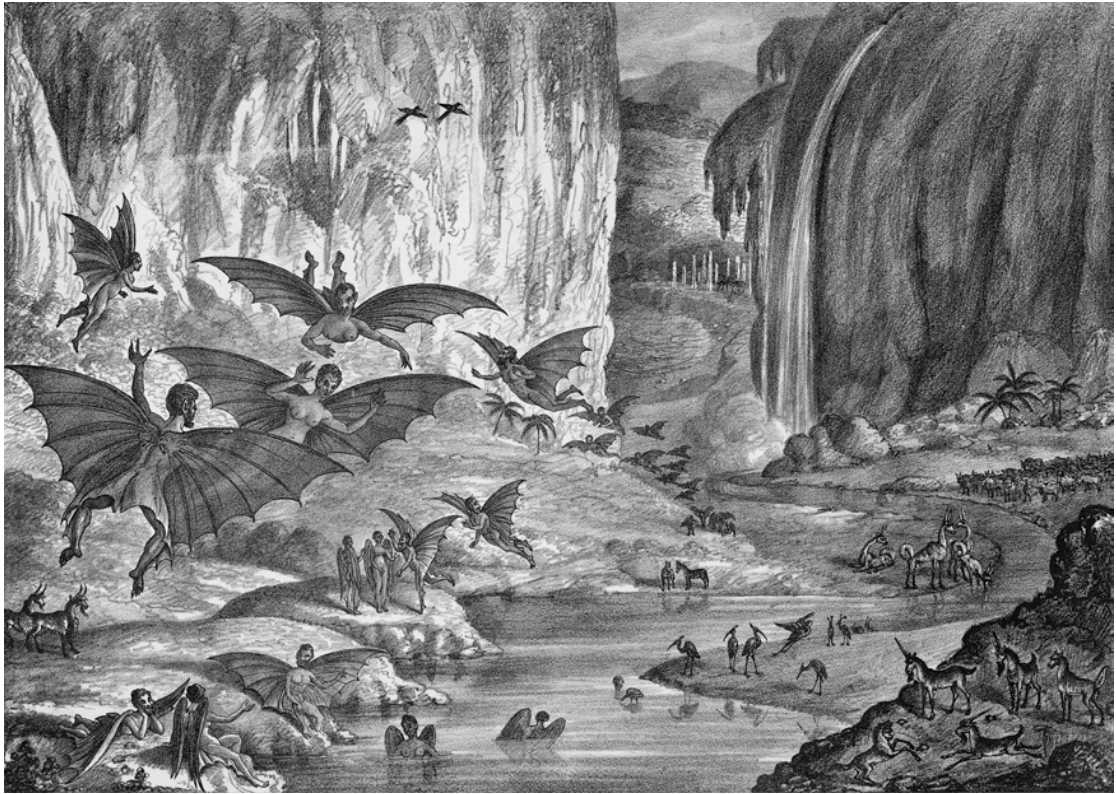

Fig. 1. A scene from the Moon, based on the descriptions published in *The Sun*, in a lithograph by Benjamin Henry Day. Source: Wikimedia Commons, public domain.

A common theme here is that extraterrestrial life was imagined as being very similar to life on Earth. This seems a priori unlikely, and it reflects the limitations of human imagination that draws on the terrestrial experience and assumes its universal validity. With a few notable exceptions, this applies to much of science fiction in a variety of media to the present day.

 *Imagining the Alien*

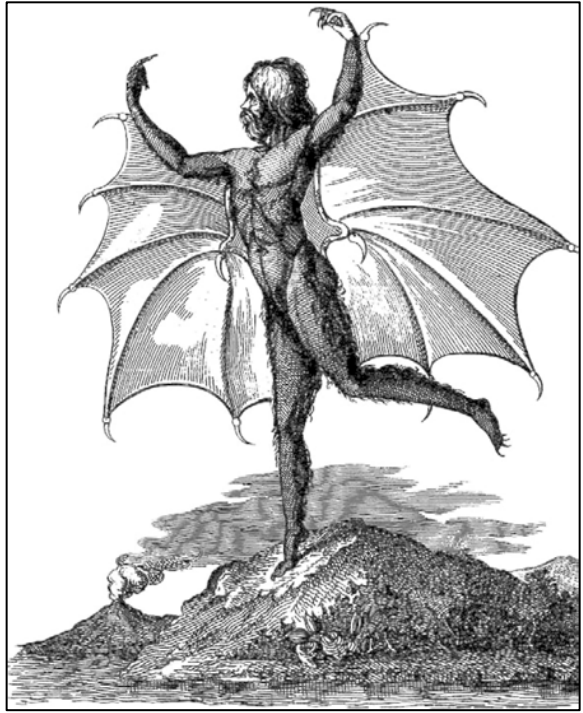

Fig. 2.  A lunar man-bat, based on the descriptions published by *The Sun*, illustrated by Leopoldo Galluzzo, in Napoli (1836).  Notice a Vesuvius-like volcano in the background. Source: Wikimedia Commons, public domain.

The next episode in this saga was the pseudo-discovery of canals on Mars[6] by the astronomer Giovanni Schiaparelli in 1877.  He thought that he could discern a network of lines on the surface of Mars and called them "channels," which was then mistranslated to English as "canals" and fueled the speculation that they are a vast irrigation system built by an advanced civilization on Mars.  We now know that these Martian canals were an optical illusion.  Interestingly, many other observers who followed Schiaparelli, including the well-known astronomer William Pickering, also thought that they could see them, perhaps illustrating the human tendency to see what they wanted to see.

The foremost among them was the astronomer Percival Lowell.  He founded the eponymous observatory in Arizona, which remains active to the present day.  Lowell completely bought into the story, including the idea that these are giant irrigation canals. He observed Mars extensively, drew many maps showing the purported canals and published three books on the subject. Alas, the canals and the putative Martian civilization that built them were an illusion, as shown conclusively by the Mariner missions in the 1960s and numerous others since then. According to Sharps (2018)[7], this may have been a result of many factors, both psychological and sociocognitive.

Lowell's true legacy is the observatory he founded, and the science done there, including the discovery of Pluto by Clyde Tombaugh in 1930.

Nevertheless, the idea of an advanced civilization on Mars was lodged in the collective human consciousness. Perhaps the best example is H. G. Wells's early science fiction novel *The War of the Worlds* (1897), which he intended as a commentary on the theory of evolution, imperialism, and Victorian-era fears, superstitions, and prejudices[8]. What really brought it to wide public attention was the famous radio play by Orson Welles (1938)[9] that created a short-lived panic in the eastern US and inspired multiple movies, books, etc. It also started an entire alien invasion science fiction genre that continues unabated.

This was an excellent example of a projection of human fears, history, and experiences on an alien canvas: the innate human fear of the alien, wars of conquest, colonialism, slavery, etc. These themes have continued through much of science fiction and other aspects of popular culture ever since, in a broad range of media, from print through the movies, television, comics, and the internet. Martians, in particular, usually played the role of aggressors, perhaps because they were from the planet named after the Greco-Roman god of war. One has to wonder what the invaders would be like if they came from Venus.

As the 20th century progressed, as societal inhibitions started to decline, the alien themes broadened to include sexual and suggestive, mainly of the 'damsel in distress' variety, often with male astronaut knights coming to their rescue. A quick perusal of pulp fiction covers from the 1930s to 1950s (easily found on the internet, but not reproducible in a respectable cultural journal) provides good and often entertaining illustrations of this phenomenon. For some reason, alien monsters (and they were always monsters) seemed to be remarkably interested in the human females, usually of a scantily clad and buxom variety. Or, perhaps, it was their intended male teenage audience.

---

Movies and television featuring aliens, as one of the dominant entertainment media over the past century, have largely shown similar human cultural and psychological predispositions. Sometimes, even the fears preceding human culture. One frequent theme is exemplified in the *Alien* franchise: there is a direct line between a sabretooth tiger in a cave and a murderous alien in a spaceship. Perhaps the most blatant compendium of the cinematic rehash of human history, fables, and just about any cliché known to Hollywood is the *Star Wars* franchise, with princesses, sword fights, cowboys, toys, Nazis, etc., set in a supposed space opera setting. Perhaps their popularity is due to the human desire for the familiar, the same psychology behind the popularity of fast-food chains.

There are, of course, some better examples that simply use science fiction settings, including aliens, to explore human personal, societal, and political issues of the day. They include the *Star Trek* franchise and several other examples dating from the Cold War era, e.g., *The Day the Earth Stood Still* (1951), *This Island Earth* (1955), and many others[10].

In the written science fiction domain, several writers made efforts to go beyond the terrestrial experiences in imagining alien encounters. They include Arthur C. Clarke, Stanislaw Lem, Ted Chiang, and a few others, who wrote about extraterrestrial life and encounters that are like nothing that we know and maybe simply incomprehensible to humans.

It is interesting to examine the evolution of Clarke's thinking about such matters. In *Childhood's End* (1953), aliens arrive and bring gifts of advanced technologies to humans but want something in return, typical of science fiction of the day. In *2001: A Space Odyssey* (1968, both the novel and the movie produced and directed by Stanley Kubrick), alien intelligence – which is never seen, aside from the mysterious black monoliths – intervenes to advance human evolution and does not want anything in return. In *Rendezvous with Rama* (1973), an alien spaceship passes through the Solar System and completely ignores human attempts to communicate with it.

In *Solaris* (1961), Stanisław Lem posits an alien intelligence in a formative stage, as a planet-spanning ocean. In *His Master's Voice* (1968), an alien message is received that describes a highly advanced technology, but one that intrinsically cannot be used for destructive purposes, unlike every technology devised by humans so far. In *Roadside Picnic* (1972) by Arkady & Boris Strugatsky, aliens arrive on the Earth, ignore humans, and depart, leaving on the landing sites bits and pieces of advanced technologies that humans try to reverse-engineer – just as ants would deal with crumbs of food left by

---

humans after a roadside picnic.  These works explore the idea that aliens are truly alien and not interested in humans one way or another.

There are obvious similarities between human imagining of alien encounters and religions: both aliens and gods inhabit the heavens, are human or humanoid in appearance and behavior, and are enormously powerful through some advanced technology or the ability to perform miracles.  This is captured by Arthur Clarke's third law, "any sufficiently advanced technology is indistinguishable from magic."  Above all, imagined as being very much like humans, both gods and aliens are interested in and meddle in human affairs.  However, as noted above, it is entirely plausible that advanced extraterrestrial intelligences may have little or no interest in humans, have forms unlike anything on this planet, and may be simply incomprehensible to us.

The connection between religion and science fiction was made explicitly by the author L. Ron Hubbard, who started Scientology, a novel religion with an origin story that – perhaps not accidentally – reads like poor science fiction from the 1950s, complete with atomic bombs and the Cold War overtones.  But to be fair to Scientology, its teachings are no more fantastic than those of some other, well-established religions, and based on approximately the same amount of evidence. Arguing which one of them is the one true religion is beyond the scope of this paper and is left to the appropriate experts.

The phenomenon of UFOs (Unidentified Flying Objects), now renamed UAPs (Unexplained Aerial Phenomena), also captures a lot of human beliefs, fears, and desires. It is probably not an accident that it really took off after World War II, which established aviation as a major and ubiquitous technology, and during the Cold War, which introduced rockets and space travel.  Sometimes, aliens arriving in these putative spaceships are seen as angels descending to the Earth to bring some great benefits to humanity.  Another psychological projection is the belief in alien abductions, whereby humans are brought into the spaceships and examined, including probing of their bodily orifices; one wonders what Sigmund Freud would make out of it.

There is copious literature about the UFO/UAP phenomena and their connection with human psychology and culture.  UFO/UAP beliefs may be a modern version of the beliefs in witches, demons, elves, etc., from the bygone times.  If there is any doubt about the cultural origins of the purported UFO sightings, Figure 3 speaks for itself.  It is of course not impossible that some UFO/UAP encounters represent actual alien probes, but so far there is no compelling evidence for their extraterrestrial origin.

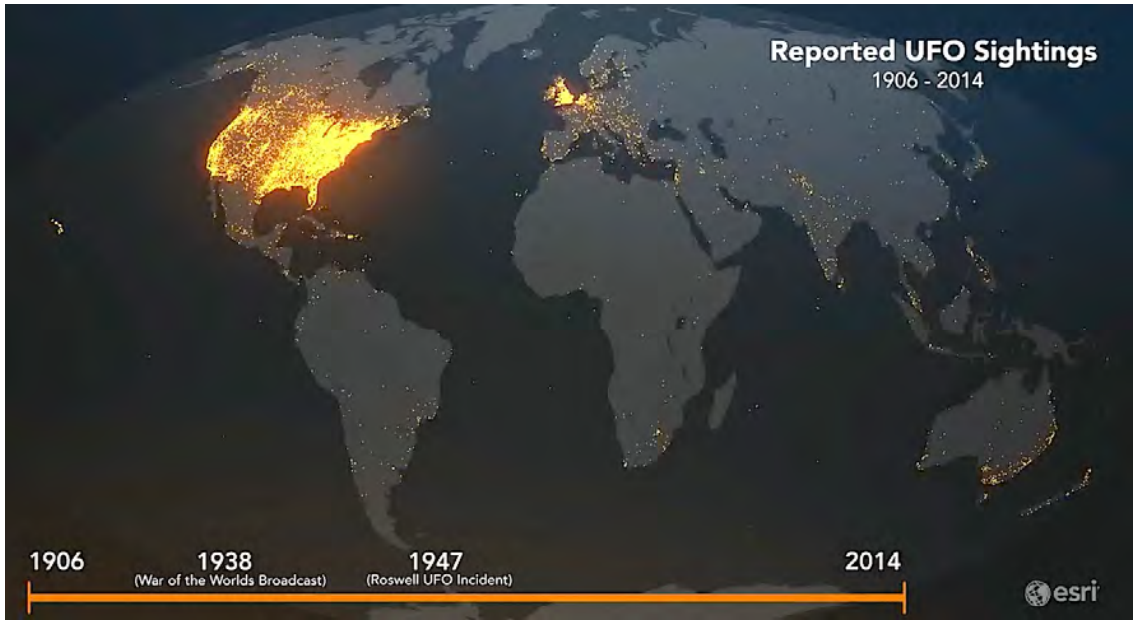

Fig. 3. The map of UFO sightings worldwide, from a video[11] produced by the cartography company Esri, based on the data provided by the National UFO Reporting Center.  The extraterrestrial visitors seem to be remarkably interested in the US and UK, but much less so in the rest of the world.

This brings up the subject of the Fermi paradox[12], the apparent conflict between the likelihood of an advanced civilization exploring our entire Galaxy in time that is much shorter than the age of the Solar system and the apparent absence of any alien technology around us.  While over a hundred possible explanations have been advanced, we note two that are along the themes of this paper. First, such advanced technology manifestation may be simply unrecognizable by human minds (say, like a group of primates somewhere in Africa, and a passenger jet flying overhead).  Second, advanced civilizations may not be interested in conquering the Galaxy in a way that mimics the European conquest of the world during the geographical age of discovery.  Again, we are projecting human experience to supposedly vastly more evolved minds.

We now know that planets are common in the universe; as of this writing, over 5000 have been confirmed, and we estimate that there are a few hundred billion in the Milky Way alone.  Some fraction of them is sufficiently Earth-like that they may support life that could be similar to the life that evolved on our planet. Thus, imagining what extraterrestrial life and intelligence may be like is a subject worthy of serious study.

---

[11] The video is available at  https://www.youtube.com/watch?v=lAopNJMbFEI

[12] Duncan Forgan, Solving Fermi's Paradox (Cambridge, UK: Cambridge Univ. Press, 2019). For an excellent summary, see https://en.wikipedia.org/wiki/Fermi_paradox

Scientifically, and in the spirit of the Copernican principle, it is perfectly reasonable to propose that life has originated on many planets and that it has evolved as it did on Earth, resulting in intelligence and technology.

Thus, Search for ExtraTerrestrial Intelligence (SETI), mostly in the form of radio signals, became a legitimate scientific undertaking, starting with the pioneering paper by Cocconi & Morrison (1959)[13] and the first actual search by Drake (Project OZMA, 1959-60), described by Drake (1961)[14]. Such searches, but with vastly more powerful radio astronomy tools, continue through the present day. There is copious literature on the subject of SETI, which we cannot review here; a great resource is the compilation of references by the Penn State Extraterrestrial Intelligence Center, maintained by Jason Wright[15]. Our point here is that what used to be purely in the realm of speculation is now a legitimate scientific subject.

However, all of the radio-based SETI searches, as well as those based on the modulated laser pulses (Optical SETI), are based on several non-trivial assumptions beyond the basic one that advanced extraterrestrial civilizations do exist: that they would want to announce their presence, that they would do this with the mid-20th century planet Earth technology, and do it in the ways that we can understand. This seems a priori highly unlikely to this author, but it does not diminish the efforts of many SETI researchers who do what they can with the tools they have. Even if these traditional SETI searches do not succeed, they still have a symbolic and cultural value.

In this regard, it is worth noting that with the birth of radio technology in the early 20th century, the three great inventors at the time, Nikola Tesla, Thomas Edison, and Guglielmo Marconi, all thought that they detected radio signals from Mars; Marconi even thought that the Martians are sending us the letter 'M' in the Morse code[16]. They were studying radio waves at frequencies of a few kHz, which we now know cannot pass through the Earth's ionosphere, but these great inventors were not aware of its existence at the time. Aliens are always assumed to want to communicate with us using the terrestrial technology of the day.

A more rigorous approach assumes that only mathematics, physics, chemistry, and information theory may be truly universal. The evolution of matter to ever greater

---

[13] Giuseppe Cocconi and Philip Morrison, 'Searching for Interstellar Communications', *Nature*, Vol. 184 (4690), pp. 844-846 (1959)

[14] Frank Drake, 'Project Ozma', *Physics Today*, Vol. 14 (4), pp. 40–46 (1961)

[15] Penn State Extraterrestrial Intelligence Center, https://www.pseti.psu.edu/

[16] Nanette South Clark, *Hello, Earth! Early 20th Century Scientists Discuss Communication with Mars*, https://anengineersaspect.blogspot.com/2013/03/hello-Earth-early-20th-century.html

complexity, reaching the emergence of intelligence, may be a natural and ubiquitous phenomenon, although its evolution may be vastly different from what happened on our planet. On the other hand, we know from the convergent evolution in terrestrial biology that very similar, complex forms and behaviors can emerge through very different evolutionary paths, although in a shared terrestrial environment, which must play a significant determinative role

Thus, a more general, modern approach to SETI is to search for possible technosignatures of advanced civilizations – large-scale manifestations of a hypothetical highly advanced technology that may be distinguishable from natural phenomena in some way.

Living organisms and civilizations consume energy and generate entropy, typically in the form of waste heat. They are entropy sources, as well as information sources. It would make sense to search for technosignatures in the form of waste heat, which is an inevitable consequence of the processing of energy. This is the idea behind Dyson spheres, hypothetical megastructures built around stars that could capture most of their luminosity, as suggested by Dyson (1960)[17]. However, many astrophysical, natural objects, e.g., black holes, dust-obscured stars, etc., are also sources of thermal radiation, and the challenge is to distinguish between natural and artificial sources.

A more objective approach to searches for technosignatures that avoids or at least minimizes human cultural, anthropocentric, and technological biases is enabled by big data. Astronomy now generates vast amounts of data, currently measured in Petabytes, over a full range of wavelengths, spanning the entire sky, detecting billions of sources, and monitoring them for changes in brightness or position. Exploration of such data using objective Machine Learning techniques can enable more objective, complete, and less biased approaches to searches for extraterrestrial technosignatures[18], as suggested by Djorgovski (2000). A report by Lazio et al. (2023)[19] provides a comprehensive discussion of this subject.

---

[17] Freeman Dyson, 'Search for Artificial Stellar Sources of Infrared Radiation', *Science*, Vol. 131 (3414), pp. 1667–1668 (1960)

[18] S. G. Djorgovski, 'Generalized SETI in a Virtual Observatory', in: *Bioastronomy 1999*, eds. K. Meech and G. Lemarchand, A.S.P. Conf. Ser. Vol. 213, pp. 519-522 (2000)

[19] Joe Lazio, S. G. Djorgovski, A. Howard, C. Cutler, S. Sheikh, S. Cavuoti, D. Herzing, K. Wagstaff, J. Wright, V. Gajjar, K. Hand, U. Rebbapragada, B. Allen, E. Cartmill, J. Foster, D. Gelino, M. Graham, G. Longo, A. Mahabal, L. Pachter, V. Ravi, G. Sussman, "Data-Driven Approaches to Searches for the Technosignatures of Advanced Civilizations", Keck Institute for Space Studies workshop report, (2023); https://doi.org/10.26206/gvmj-sn65 , https://arxiv.org/abs/2308.15518

Finally, intelligence need not be biological in origin. We may have created an effectively alien intelligence here, on planet Earth, in the form of Artificial Intelligence (AI). AI is based on vastly different principles, materials, and architectures than biological intelligence. It currently evolves on time scales that are about a million times faster than biological evolution, and it is only a matter of time before it surpasses human intelligence. Not surprisingly, it has also triggered the same fears and projections in the popular culture as putative aliens did in the past: AI is the new Martians.

It is impossible to predict future developments as biological and non-biological intelligences co-exist on this planet. Exponential biological evolution of complexity and intelligence of organisms gave rise to a much faster technological evolution of non-biological intelligence. What happened on Earth may have happened elsewhere in the universe as well. We may learn from AI what to look for and how to search for aliens, or maybe something even more interesting.

Even if there is advanced intelligence of an extraterrestrial equivalent of biological origin, their goals, behavior, and technology may be beyond our intellectual capabilities. Our household pets (as well as primates, cetaceans, etc.) do have a certain degree of intelligence but are incapable of understanding much of the human culture or science, even though we are the products of the same evolutionary process in the same planetary environment. We may be in a similar position relative to the advanced extraterrestrial minds.

To conclude, imagining of extraterrestrial alien beings and civilizations throughout history holds a metaphorical mirror of human culture, fears, values, and psychology, but it does not say much about what these alien entities may be really like. At first glance, this seems like a colossal failure of our individual and collective imagination, but it may be a result of deeper cognitive limitations of human minds. Our thinking and culture are shaped by our environment and the experiences we have in interacting with it. It is certainly very hard, and perhaps impossible, to transcend these limitations and imagine what beings that evolve in completely different planetary contexts may be like. But maybe we'll find out someday.